\documentclass[12pt]{iopart}
\usepackage{iopams}

\usepackage{tabularx}

\usepackage[final]{graphicx}% Include figure files
\usepackage{dcolumn}% Align table columns on decimal point
\usepackage{bm}% bold math

\usepackage{longtable}
\usepackage{tabularx}

\def\extgra{pdf}

\newcommand{\mm}[1]     {\ifmmode {#1} \else{}${#1}$\fi}
\newcommand{\mmm}[1]    {\ifmmode{}#1 \else{}${#1}$\fi}
\newcommand{\beq}[1]    {\begin{equation} \label{#1}}
\newcommand{ \eeq}{\end{equation}}

\def\fesex{\mm{\rm{ Fe_{2-x}Se_2Cs_{y} } } }
\def\fesex{\mm{\rm{ Cs_{y}Fe_{2-x}Se_2 } } }
\def\fesexx{\mm{\rm{ X_{y}Fe_{2-x}Se_2 } } }
\def\fesetl{\mm{\rm{ Tl_{y}Fe_{2-x}Se_2 } } }
\def\fesek{\mm{\rm{ K_{y}Fe_{2-x}Se_2 } } }
\def\feserb{\mm{\rm{ Rb_{y}Fe_{2-x}Se_2 } } }

\def\fese{\mm{\rm{ FeSe } } }

\def\Tc\mm{T_c}

\begin{document}


\title[Room
temperature AFM order in superconducting $\rm{X_{y}Fe_{2-x}Se_2}$, $\rm{( X= Rb, K)}$] {Room
temperature antiferromagnetic order in superconducting $\rm\mathbf{X_{y}Fe_{2-x}Se_2}$, $\rm\mathbf{( X= Rb, K)}$: a powder neutron diffraction study.}


\author{V.~Yu.~Pomjakushin$^{1}$, E.~V.~Pomjakushina$^{2}$, A.~Krzton-Maziopa$^{2}$, K.~Conder$^{2}$, Z.~Shermadini$^{3}$ }
\address{$^{1}$ Laboratory for Neutron Scattering, Paul
Scherrer Institut, CH-5232
Villigen PSI, Switzerland}

\address{$^{2}$ Laboratory for Developments and Methods, PSI, CH-5232
Villigen PSI, Switzerland}

\address{$^{3}$ Laboratory for Muon Spin Spectroscopy, Paul
Scherrer Institut, CH-5232 Villigen PSI, Switzerland}

\ead{Vladimir.Pomjakushin@psi.ch}
%\pagebreak

%\date{\today}

\begin{abstract}

Magnetic and crystal structures of superconducting \fesexx\ (X= Rb and K with $T_c=$31.5~K and 29.5~K) have been studied by neutron powder diffraction at room temperature. Both crystals show ordered iron vacancy pattern and the crystal structure is well described in the $I4/m$ space group with the lattice constants $a=$8.799, $c=$14.576 and $a=$8.730, $c=$14.115~\AA\, and the refined stoichiometry x=0.30(1), y=0.83(2) and x=0.34(1), y=0.83(1) for Rb- and K-crystals respectively. The structure contains one fully occupied iron position and one almost empty vacancy position. Assuming that the iron moment is ordered only on the fully occupied site we have sorted out all eight irreducible representations (irreps) for the propagation vector $k=0$ and have found that irreps $\tau_2$ and $\tau_7$ well fit the experimental data with the moments along $c$-axis. The moment amplitudes amounted to 2.15(3)~$\rm\mu_B$, 2.55(3)~$\rm\mu_B$ for $\tau_2$ and 2.08(6)~$\rm\mu_B$, 2.57(3)~$\rm\mu_B$ for $\tau_7$ for Rb- and K-crystals respectively.  Irrep $\tau_2$ corresponds to the Shubnikov group $I4/m'$ and gives a constant moment antiferromagnetic configuration, whereas $\tau_7$ does not have Shubnikov counterpart and allows two different magnetic moments in the structure. 

\end{abstract}

\pacs{75.50.Ee, 75.25.-j, 61.05.C-, 74.90.+n}

\submitto{\JPCM}

\maketitle

\section{Introduction}
The recent discovery of the Fe-based superconductors has triggered a remarkable renewed interest for possible new routes leading to high-temperature superconductivity. As observed in the cuprates, the iron-based superconductors exhibit interplay between magnetism and superconductivity suggesting the possible occurrence of unconventional superconducting states. Other common properties are the layered structure and the low carrier density. Among the iron-based superconductors \fese\ has the simplest structure with layers in which Fe cations are tetrahedrally coordinated by Se \cite{Hsu2008}. Recently superconductivity at about 30K was found in \fesexx\  for X=K, Cs, Rb \cite{PhysRevB.82.180520,Krzton2010,2010arXiv1012.5525W,2011arXiv1101.5670L}. Muon-spin rotation/relaxation ($\mu$SR) experiments evidence that the superconducting state observed in \fesex\ below 28.5(2) K is microscopically coexisting with a magnetic phase with a transition temperature at $T_m=478.5(3)$ K \cite{2011arXiv1101.1873S}. Recently the AFM order was reported in superconducting $\rm {K_{0.8}Fe_{1.6}Se_2}$ with $T_{\rm N}=560$~K with the iron magnetic moment 3.31~$\rm\mu_B$~\cite{Bao2011}  and \fesex\ with the magnetic moment about 2$\rm\mu_B$ per iron site \cite{pom2011fesex}. In the latter paper several possible symmetry adapted magnetic configurations have been proposed, including the one with non-constant moment configuration. During the preparation of the present manuscript a new paper appeared confirming the AFM order in Cs-crystals and reporting on the similar AFM structure in Rb-crystals with the moment 1.9(2)~$\rm\mu_B$ \cite{2011arXiv1102.2882Y}.  

The average crystal structure of \fesexx\ is the same as in the layered (122-type) iron pnictides with the space group $I4/mmm$ \cite{PhysRevLett.101.107006}. Different types of iron vacancy ordering in \fesetl\ were observed long time ago \cite{Sabrowsky,Haggstrom}, including the one with 5 times bigger unit cell. Due to renewed interest to the superconducting chalcogenides many new experimental studies on the vacancy ordering in \fesexx\ (X=K,Tl) have appeared very recently \cite{Fang2010,Wang2011,Zavalij2011,Wang2011,Basca2011,Bao2011}. In the previous paper \cite{pom2011fesex} we have determined the iron vacancy superstructure with k-vector star $\{[{2\over5},{1\over5},1]\}$ in \fesex\ by means of single crystal x-ray reciprocal space mapping and proposed two symmetry adapted magnetic configurations for the superstructure with $I4/m$ space group that well fitted the powder neutron diffraction data combined with the single crystal dataset.  The coexistance of the superconductivity and the magnetic order with an extraordinary high N\'{e}el temperature is very unusual and the direct proofs of the presence of the long range AFM ordering and the determination of the magnetic configurations by means of neutron diffraction seem to be very important. In this paper we report on the room temperature crystal and magnetic structures in the vacancy ordered superconducting \fesexx\ (X=Rb, K).

\section{Samples. Experimental}
\label{exp}

Single crystals of intercalated iron selenides of nominal compositions $\rm {X_{0.8}(FeSe_{0.98}})_2$  were grown from the melt using the Bridgman method as described in Ref.~\cite{Krzton2010}. The superconducting transition has been detected by ac-susceptibility using a conventional magnetometer. The sample holder contains a standard coil system with a primary excitation coil (1300 windings, 40 mm long) and two counter-wound pick-up coils (reference and sample coil, each 10 mm long and 430 windings) which are connected to a lock-in amplifier. The frequency used was 144 Hz and the sample holder diameter was 5 mm. Measurements were performed by heating the sample at a rate of 9 K/h. The onset of the critical temperature has been determined to be $T_{c}$ = 31.5 K and 29.5 K for the Rb and K intercalated compounds, respectively. The magnetometer was calibrated using the superconducting transition of a lead sample showing a 100\% superconducting fraction. The raw data were then normalized to the sample volume relative to the one of the Pb calibration specimen. The superconducting fraction determined this way amounted to 73-100\% and 22\% for the Rb- and K-crystals respectively. Neutron powder diffraction experiments were carried out at the SINQ spallation source of Paul Scherrer Institute (Switzerland) using the high-resolution diffractometer for thermal neutrons HRPT \cite{hrpt} ($\lambda=1.866, 1.494$~\AA, high intensity mode $\Delta d/d\geq1.8\cdot10^{-3}$) at room temperature. For the powder diffraction measurements pieces of corresponding crystal were powdered and loaded into a vanadium container with an indium seal in an He glove box. Inasmuch as K-containing sample was measured just after the first successful synthesis, we were not aware, that the tip of crystal taken form Bridgman ampoule always contain impurity phases (predominantly Fe) and we have measured mixture of precipitated impurities together with powdered crystal. As for Rb-crystal we have powdered only a middle part of the crystal, which looked a single domain with shiny surface. Refinement of crystal and magnetic structures of powder neutron diffraction data were carried out with {\tt FULLPROF}~\cite{Fullprof} program, with the use of its internal tables for scattering lengths and magnetic form factors.

\section{Results and discussion}
\label{res}

\begin{figure}[b]
  \begin{center}
    \includegraphics[width=10cm]{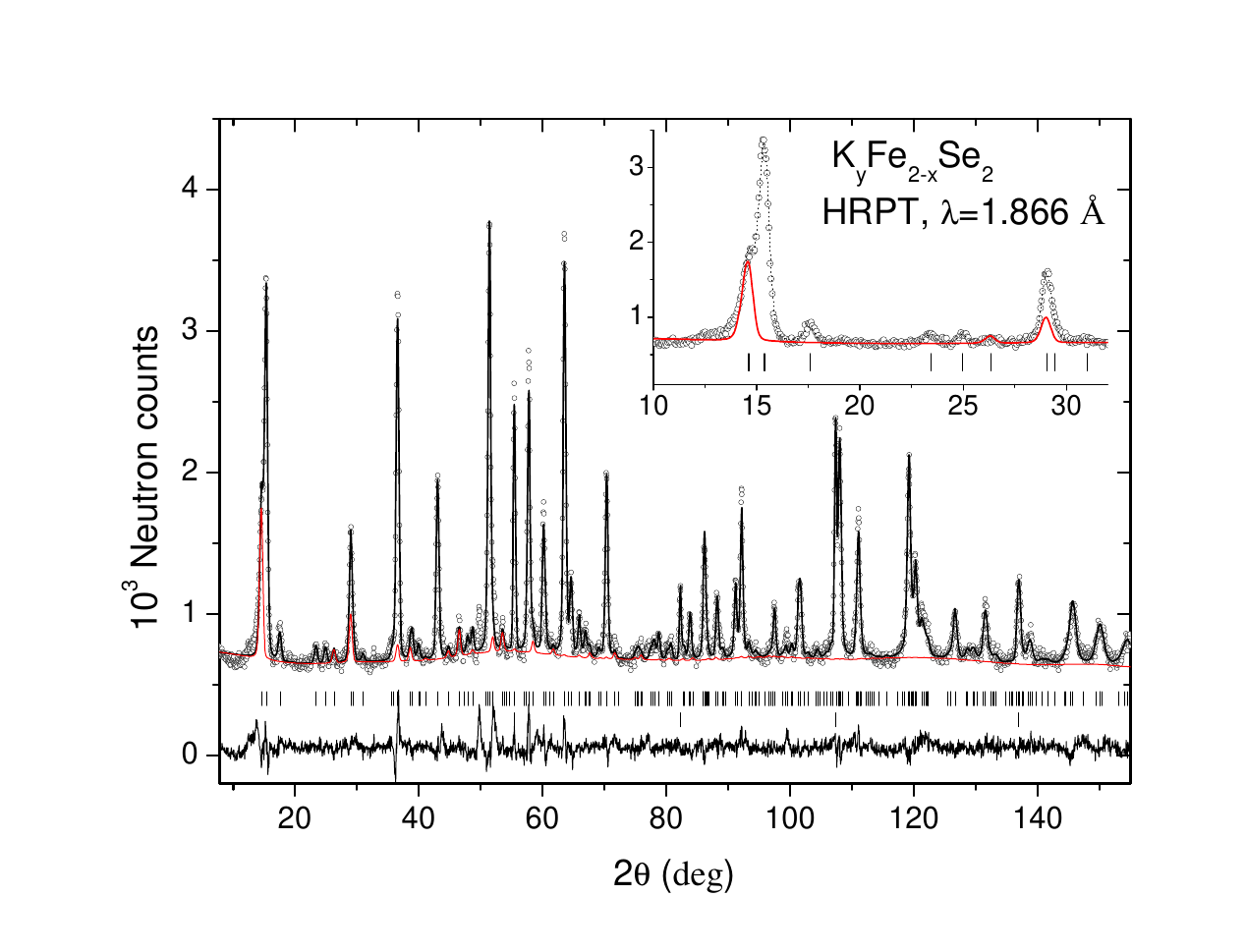} %fesex_mag

  \end{center}

\caption{
The Rietveld refinement pattern and difference plot of the neutron diffraction data for \fesek\ at room temperature (T=300K) measured at HRPT with the wavelength $\lambda=1.886$~\AA. The rows of tics show the Bragg peak positions for the main phase and Fe-impurity.  The magnetic contribution together with the background is shown by red line. The inset shows the zoomed low two theta domain.
}
 \label{fig1}
\end{figure}

\begin{figure}[b]
  \begin{center}
    \includegraphics[width=10cm]{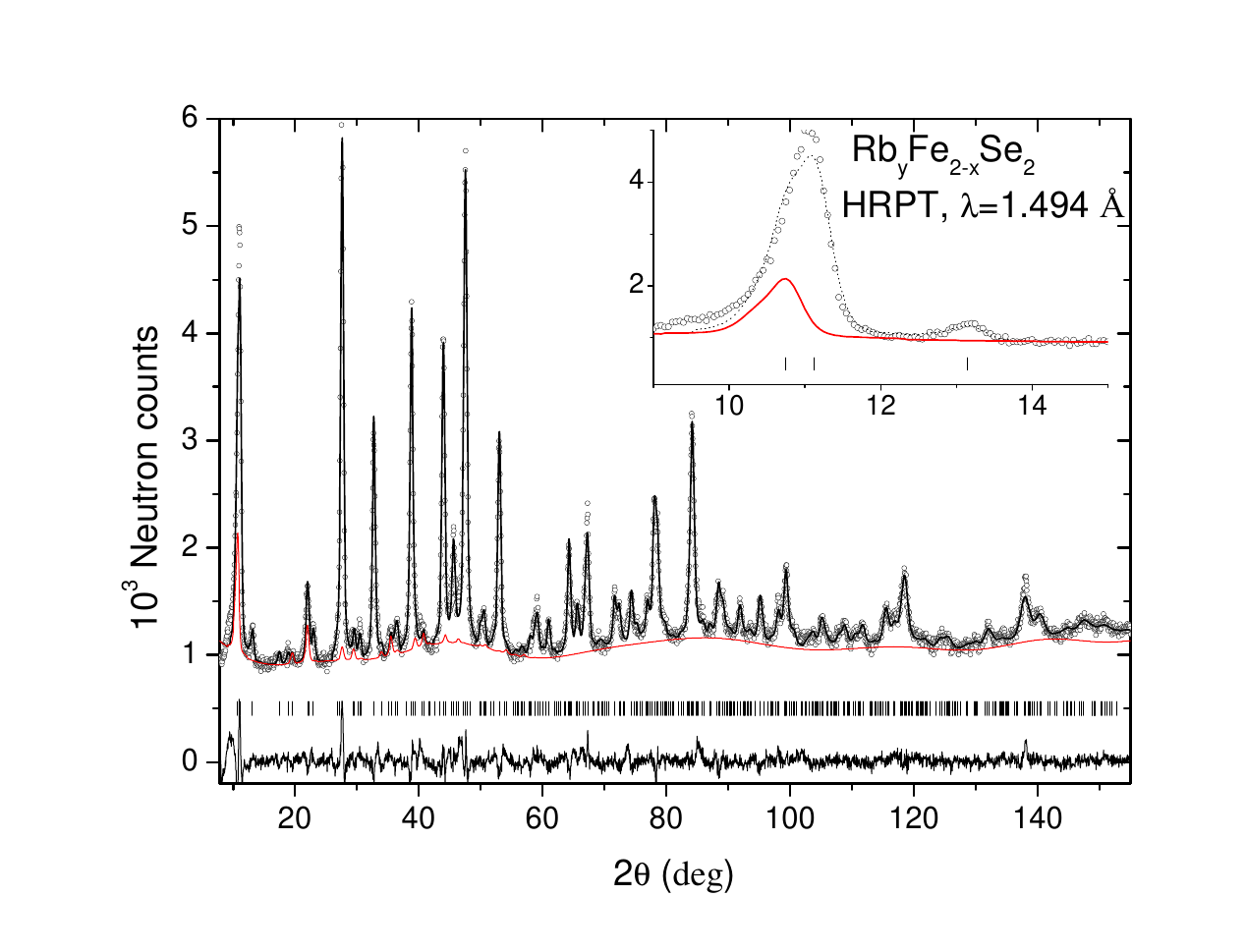} %fesex_mag
  \end{center}
\caption{
The Rietveld refinement pattern and difference plot of the neutron diffraction data for \feserb\ at room temperature (T=300K) measured at HRPT with the wavelength $\lambda=1.494$~\AA. The rows of tics show the Bragg peak positions. The magnetic contribution together with the background is shown by red line. The inset shows the zoomed low two theta domain with the magnetic contribution indicated by the red line.}
  \label{fig2}
\end{figure}

\begin{figure}
  \begin{center}
    \includegraphics[width=8cm]{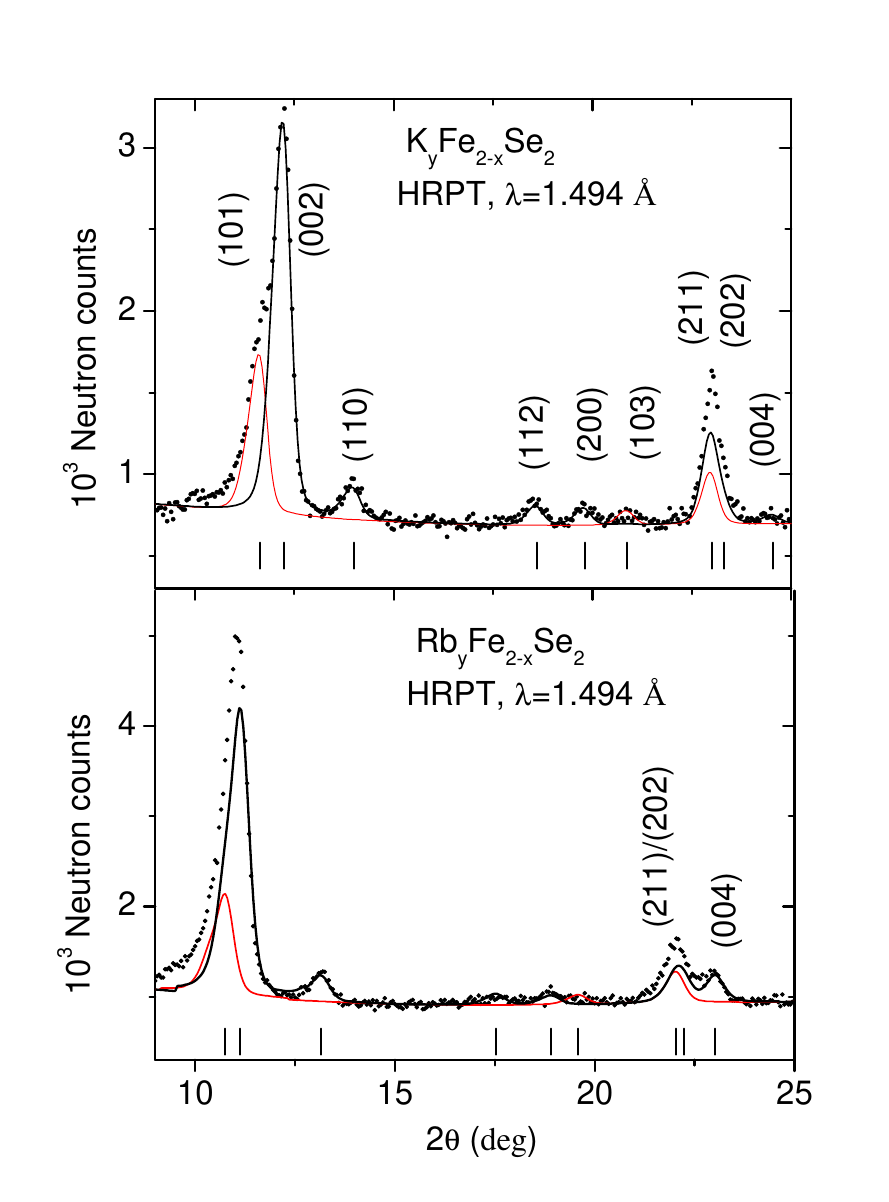} %fesex_mag
  \end{center}
\caption{Fragments of NPD patterns for \fesexx\ (X=K, Rb) at room temperature for $\lambda=1.494 {\rm\,\, \AA}$ and the partial magnetic and nuclear contributions shown by red and black lines. The rows of tics show the Bragg peak positions.}
  \label{fig3}
\end{figure}

\begin{figure}
  \begin{center}
    \includegraphics[width=9cm]{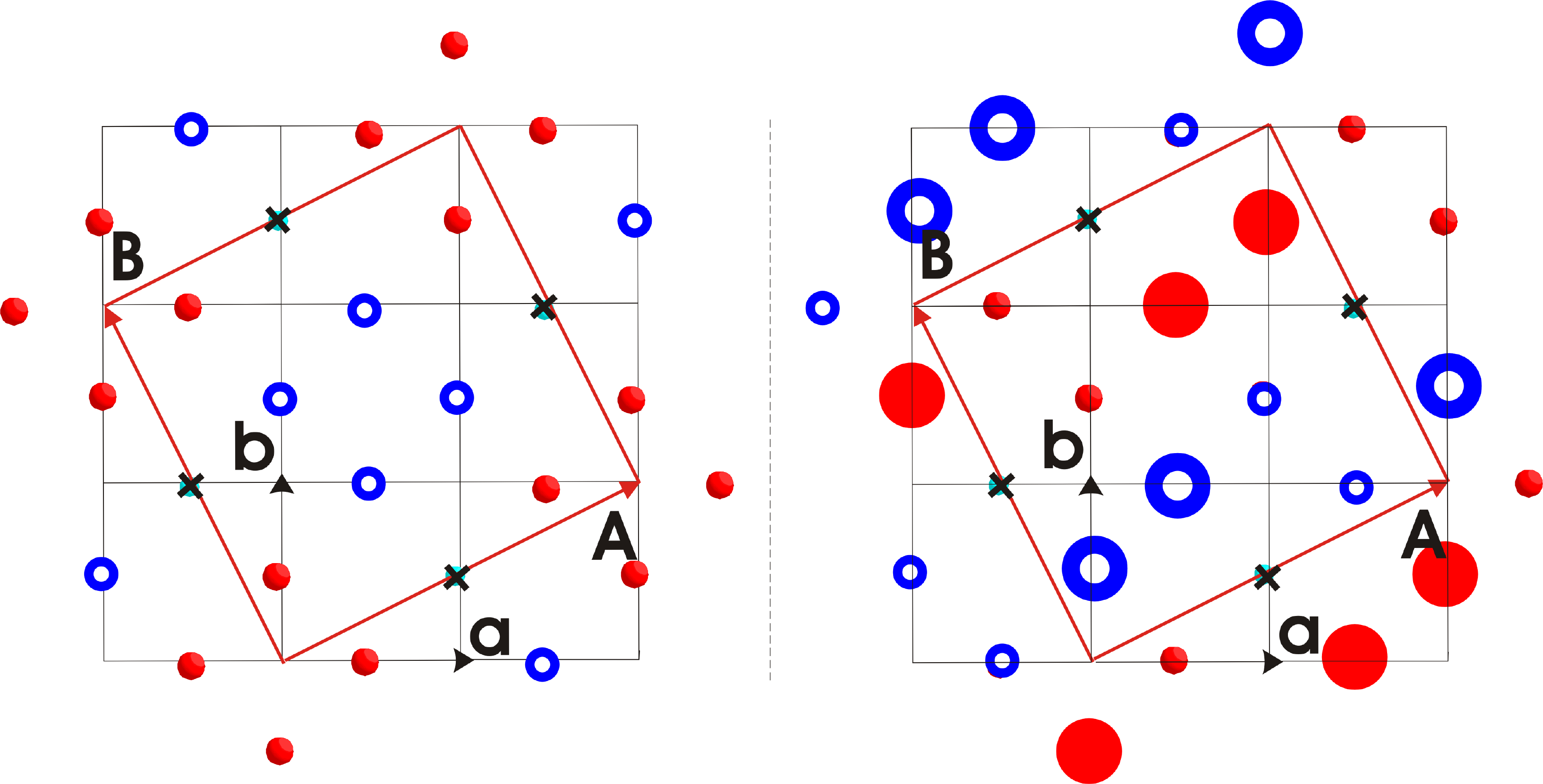} %fesex_mag
  \end{center}
\caption{Magnetic configurations for $\tau_2$ (left) and $\tau_5$ (right) irreps with the Fe2 magnetic moments along $c$-axis.  Figure shows a projection of one layer of Fe on $(ab)$-plane. The neighboring planes are arranged antiparallel. Fe1 vacancy positions are shown by the crossed cyan circles. $\tau_5$-structure can have two different moment sizes indicated by large and small circles. Open blue and filled red circles show Fe2 down and up spins. Big red unit cell ($\mathbf{A,B}$) corresponds to $I4/m$ cell, the smaller cells ($\mathbf{a,b}$) are the cells of the average vacancy-disordered $I4/mmm$  (122)-structure.} 
  \label{magconf}
\end{figure}

Figure \ref{fig1} shows the neutron powder diffraction pattern (NPD) and the calculated profile for \fesek. The refinement of the data were performed using two NPD datasets collected at different wavelengths $\lambda=1.866$~\AA\ and $1.494$~\AA. This allows one to have better resolution for the magnetic reflections from longer lambda and better determination of the crystal structure parameters from the shorter wavelength. Since both magnetic and nuclear scattering contribute to the same Bragg peaks it is important to have big enough q-range to refine both contributions simultaneously. Figure \ref{fig2} shows NPD pattern and its refinement for the \feserb\ sample. For the crystal structure we use the vacancy ordered superstructure model $I4/m$ stated in Ref. \cite{pom2011fesex}, for the magnetic structure analysis we use symmetry adapted basis functions of eight irreducible representations (irreps) of $I4/m$ space group and propagation vector $k=0$ \cite{pom2011fesex}. The description of the structure model and the magnetic and structure parameters and the details of the refinements are summarized in Table \ref{T1}. Interestingly that in \fesex\ the lattice constants $a$ and $c$ have the ratio $(c/a)^2\simeq3$ resulting in a complete overlap of the most important Bragg peaks with the magnetic and nuclear contribution and disentangling the magnetic and crystal structures were possible only using both x-ray and neutron data \cite{pom2011fesex}. For \fesexx (X=K, Rb) this is not the case as illustrated in Fig.~\ref{fig3}. The diffraction peaks with the indicies groups (101)/(002), (200)/(103) and (211)/201)/(004) are quite well resolved, whereas in Cs-case they were completely overlapped. Similar to the analysis of \fesex\ we have sorted out all irreps (Table 3 of Ref.~\cite{pom2011fesex}) and found that the best fit of our data is achieved for the $\tau_2$ and $\tau_5/\tau_7$ irreps for the basis vectors parallel to $c$-axis (similar as we have found before for \fesex). The configuration for $\tau_7$ is equivalent to $\tau_5$ after $\pi/2$ rotation around $c$-axis. Due to better resolution of the magnetic and nuclear contributions and the better statistics neutron data sets in comparison with \fesex\ we are quite confident in the magnetic configurations that we propose. The basis vectors are explicitly listed in Ref.~\cite{pom2011fesex} and the two magnetic configurations are graphically illustrated in Fig.~\ref{magconf}. The structures with the spins laying in the (ab) plane give considerably worth reliability factors. For comparison we show the refinement details for $\tau_4$ with the moments in the $(ab)$-plane that gives better R-factors among the other ``bad'' magnetic models. Except worse overall reliability factors the $\tau_4$ structure predicts substantial intensity in the Bragg beak (103), but experimentally observed intensity is close to zero as shown in Fig.~\ref{fig3}. The magnetic model $\tau_2$ corresponds to the Shubnikov magnetic group $I/4m'$ and gives constant moment configuration as shown in Fig.~\ref{magconf}, whereas complex $\tau_7/\tau_5$ irreps with Herring coefficient 0 does not have Shubnikov counterpart. In addition, $\tau_7$ gives in general a non-constant moment configuration, as shown in Fig.~\ref{magconf}. The basis function $\psi_7$ (Table 3 of Ref.~\cite{pom2011fesex}) can be multiplied by an arbitrary phase factor $\exp(i\varphi)$. The overall phase $\varphi$ does not affect the magnetic Bragg peak intensity at all, because it is proportional to the square of absolute value of the structure factor $|F|^2$ for unpolarized neutrons. However, the iron moments marked by large and small circles shown in Fig.~\ref{magconf} will depend on  $\varphi$-value as $m\cos(\varphi)$ and $m\sin(\varphi)$, where m is the value from Table \ref{T1}. For the constant moment configuration with $\phi=\pi/4$, the moment will be $\sqrt{2}$ smaller than the maximal $m$-value. In principal, the value of $\phi$ is accessible in the present $k=0$ case from the interference term in the polarized neutron diffraction experiment.

%We would like to note that the correlation between atomic crystal structure and the magnetic structure parameters

\section*{A{\lowercase{cknowledgements}}}

The authors thank the NCCR MaNEP project and Sciex-NMS$\rm ^{ch}$ (Project Code 10.048) for the support of this study. The work was partially performed at the neutron spallation source SINQ. Fruitful discussions with D.~Sheptyakov are gratefully acknowledged.

\section*{References}

\bibliography{../../refs/refs_general,../../refs/refs_fesex,../../refs/refs_manganites,../../publication_list/publist2007,../../publication_list/publist2006,../../publication_list/publist2005,../../publication_list/publist}


\newcommand{\PBS}[1]{\let\temp=\\#1\let\\=\temp}

\begin{table}
\caption{Crystal and magnetic structure parameters of \fesexx\ (X=Rb, K) refined in the space group $I4/m$ (no. 87) with the atoms in the Wyckoff positions: X1 $(0,0,0)$ (2a), X2 $(x,y,0)$ (8h), Se1 in $(x,y,z)$ (16i), Se2 $({1\over2},{1\over2},z)$ (4e), Fe1 (16i) and Fe2 $({1\over2},0,{1\over4})$ (4d). The occupancies o-X, o-Fe are calculated to be in units of the formula \fesexx. The occupancy of Fe1 was fixed to the fully occupied value, which corresponds to stoichiometry 1.6. Atomic displacement parameters $B$ ($\rm\AA^2$) were constrained to be the same for the same atom types. $\chi^2$ is the global chi-square (Bragg contribution). Reliability Bragg $R_B$-factors are given for the nuclear and magnetic phases respectively. For \fesek\ we have performed a combined refinement of two datasets measured with wavelengts $\lambda_1=1.494\,{\rm \AA}$ and $\lambda_2=1.866\,{\rm \AA}$. The magnetic moment amplitudes are in $\rm\mu_B$. The magnetic configurations for two magnetic models ($\tau_2,\tau_7$) that fit the data equally well are shown in Fig.~\ref{magconf}. $\tau_4$ fits the data considerably worse, but is given in the table for comparison. }
 \label{T1}

\begin{center}
\begin{tabular}{l
                 >{\PBS\raggedright\hspace{0pt}}p{6cm}
                 >{\PBS\raggedright\hspace{0pt}}p{6cm}}

                  &    Rb                         &    K         \\ \hline
 $a$                & 8.79962(24)                   &   8.73019(12)   \\
 $c$                & 14.57615(53)                  & 14.11485(30)\\
 $x,y$ X2           & 0.399(3) 0.803(3)             &  0.374(3) 0.816(5)\\
 o-X1             & 0.59(2)                       & 0.60(1)\\
 o-X2             & 0.24(2)                       & 0.23(1)\\
 $B$ X              & 3.9(1)                        &  2.2(3)\\
 $x,y,z$ Se1        & 0.3880(9) 0.794(1) 0.6495(3)& 0.3867(8) 0.795(1) 0.6448(3)\\
 $z$ Se2            &  0.145(1)                     & 0.142(1)\\
$B$ Se              & 1.57(6)                       &  1.43(6)\\
 $x,y,z$ Fe2        &0.2958(8)  0.5920(8) 0.2524(8) & 0.2953(8) 0.5914(6)  0.2495(9)\\
 o-Fe1            & 0.099(6)                      & 0.069(6)\\
$B$ Fe              & 1.79(6)                       &  1.55(5)\\ \hline
$\tau_2$          &                               &    \\
$m || c$ Fe2                 & 2.15(6)                        & 2.55(3) \\
$\chi^2$                     & 2.78                           & 2.18           \\
$R_{B}$,\% $\lambda_1$       & 5.36 12.7                      & 6.13 7.58        \\
$R_{B}$,\% $\lambda_2$       &                                & 5.62 10.3        \\
$\tau_7,\tau_5$   &                                &    \\
$m || c$ Fe2                 & 2.08(6)                        & 2.57(3) \\
$\chi^2$                     & 2.80                           & 2.20           \\
$R_{B}$,\% $\lambda_1$       & 5.36 13.4                      & 6.17 8.69        \\
$R_{B}$,\% $\lambda_2$       &                                & 5.64 12.2        \\
$\tau_4$   &                                &    \\
$m\perp c$ Fe2               & 1.71(6)                        & 2.46(4) \\
$\chi^2$                     & 2.92                           & 2.56           \\
$R_{B}$,\% $\lambda_1$       & 5.91 16.5                      & 6.81 19.1        \\
$R_{B}$,\% $\lambda_2$       &                                & 6.32 23.5        \\
\end{tabular}
\end{center}
\end{table}

\end{document}